\documentstyle[12pt]{article}
\def\fun#1#2{\lower3.6pt\vbox{\baselineskip0pt\lineskip.9pt
\ialign{$\mathsurround=0pt#1\hfil##\hfil$\crcr#2\crcr\sim\crcr}}}

\newcommand{\be}{\begin{eqnarray}}
\newcommand{\ee}{\end{eqnarray}}
\newcommand{\ba}{\begin{array}}
\newcommand{\ea}{\end{array}}

\newcommand{\AmS}{{\protect\the\textfont2
  A\kern-.1667em\lower.5ex\hbox{M}\kern-.125emS}}

\input{epsfig.sty}

\begin{document}

\Huge{\noindent{Istituto\\Nazionale\\Fisica\\Nucleare}}

\vspace{-3.9cm}

\Large{\rightline{Sezione SANIT\`{A}}}
\normalsize{}
\rightline{Istituto Superiore di Sanit\`{a}}
\rightline{Viale Regina Elena 299}
\rightline{I-00161 Roma, Italy}

\vspace{0.65cm}

\rightline{INFN-ISS 96/10}
\rightline{October 1996}

\vspace{1.5cm}

\begin{center}

\Large{CHARMED DECAYS OF THE B-MESON\\ IN THE QUARK MODEL}
\footnote{To appear in Nuclear Physics B Suppl.: Proceedings of the 
$2^{nd}$ International Conference on {\em Hyperons, Charm and Beauty 
Hadrons}, Montr\'eal, Qu\'ebec, Canada, 27-30 August 1996.}\\

\vspace{1cm}

\large{I.L. Grach, I.M. Narodetskii, K.A. Ter-Martirosyan}

\vspace{0.25cm}

\normalsize{Institute for Theoretical and Experimental Physics,\\
            117259 Moscow, Russia}

\vspace{0.5cm}

\large{S. Simula}

\vspace{0.25cm}

\normalsize{Istituto Nazionale di Fisica Nucleare, Sezione Sanit\`{a},\\
            Viale Regina Elena 299, I-00161 Roma, Italy}

\end{center}

\vspace{1cm}

\begin{abstract}

\noindent Exclusive and inclusive, semileptonic and non-leptonic,
charmed decays of the B-meson are investigated in the context of a 
phenomenological quark model. Bound-state effects are taken care of 
by adopting a single (model-dependent) non-perturbative wave function, 
describing the motion of the light spectator quark in the $B$-meson. 
A nice reproduction of both exclusive and inclusive semileptonic data 
is obtained. Our predictions for the electron spectrum are presented and 
compared with those of the Isgur-Scora-Grinstein-Wise quark model. Finally, 
our approach is applied to the calculation of inclusive non-leptonic widths, 
obtaining a remarkable agreement with experimental findings.
\end{abstract}

\newpage

\pagestyle{plain}

\section{INTRODUCTION}

Weak decays of hadrons containing a heavy quark could provide unique 
information on the fundamental parameters of the Standard Model and, 
at the same time, could serve as a probe of our understanding of the 
non-perturbative strong interaction phenomenology. From the theoretical 
point of view, the Heavy Quark Effective Theory ($HQET$) \cite{HQET} is 
widely recognized as a very powerful tool for investigating decay modes 
of heavy flavours and, recently \cite{OPE}, a model-independent framework 
has been developed to treat non-perturbative $QCD$ effects in inclusive 
decays. The latter approach relies on the formalism of $HQET$ and on the 
use of the operator product expansion ($OPE$) in the physical region of 
time-like momenta. Thus, the hypothesis of quark-hadron duality, in its 
global form for semileptonic ($SL$) decays and in its local form for 
non-leptonic ($NL$) processes, has to be invoked \cite{PQW76}. The 
concept of quark-hadron duality, though it has not yet been derived 
from first principles, is essential in $QCD$ phenomenology and 
corresponds to the assumption that the sum over many hadronic final 
channels eliminates bound-state effects related to the specific structure 
of each individual final hadron. The validity of the global duality has 
been tested in inclusive hadronic $\tau$ decays \cite{GN96}, whereas the 
possibility of a failure of the local duality in inclusive $NL$ processes 
has been raised recently in \cite{AMPR96}. Therefore, the use of 
phenomenological quark models for the description of the hadron structure 
could be still of interest and, in this respect, it is well known that 
the constituent quark model is remarkably successfull in describing the 
hadron mass spectra. However, any model of hadrons must go beyond the 
mass spectroscopy and predict, e.g., decay processes.

\section{OUTLINE OF THE APPROACH AND RESULTS}

In this contribution the exclusive and inclusive, $SL$ and $NL$, charmed 
decays of the B-meson are investigated within the phenomenological quark 
model of Refs. \cite{MT96,GNST}, where all the non-perturbative $QCD$ 
bound-state effects are mocked up by a (model-dependent) light-cone 
wave function $\chi(x, \vec{p}_{\perp})$, describing the internal motion 
of the light spectator-quark inside the $B$-meson (with $x$ and 
$\vec{p}_{\perp}$ being the internal light-cone variables). The model 
dependence is illustrated adopting different $B$-meson wave functions, 
namely the phenomenological one of Ref. \cite{MT96} and the two 
light-cone wave functions of Ref. \cite{GNST}, constructed from a 
relativized \cite{GI85} and a non-relativistic \cite{NCS92} constituent 
quark model. In what follows, we will refer to these wave functions as 
cases $A$, $B$ and $C$, respectively.

\begin{figure}[htb]
\begin{center}
\vspace{0.5cm}
\parbox{12cm}{\epsfxsize=12cm \epsfbox{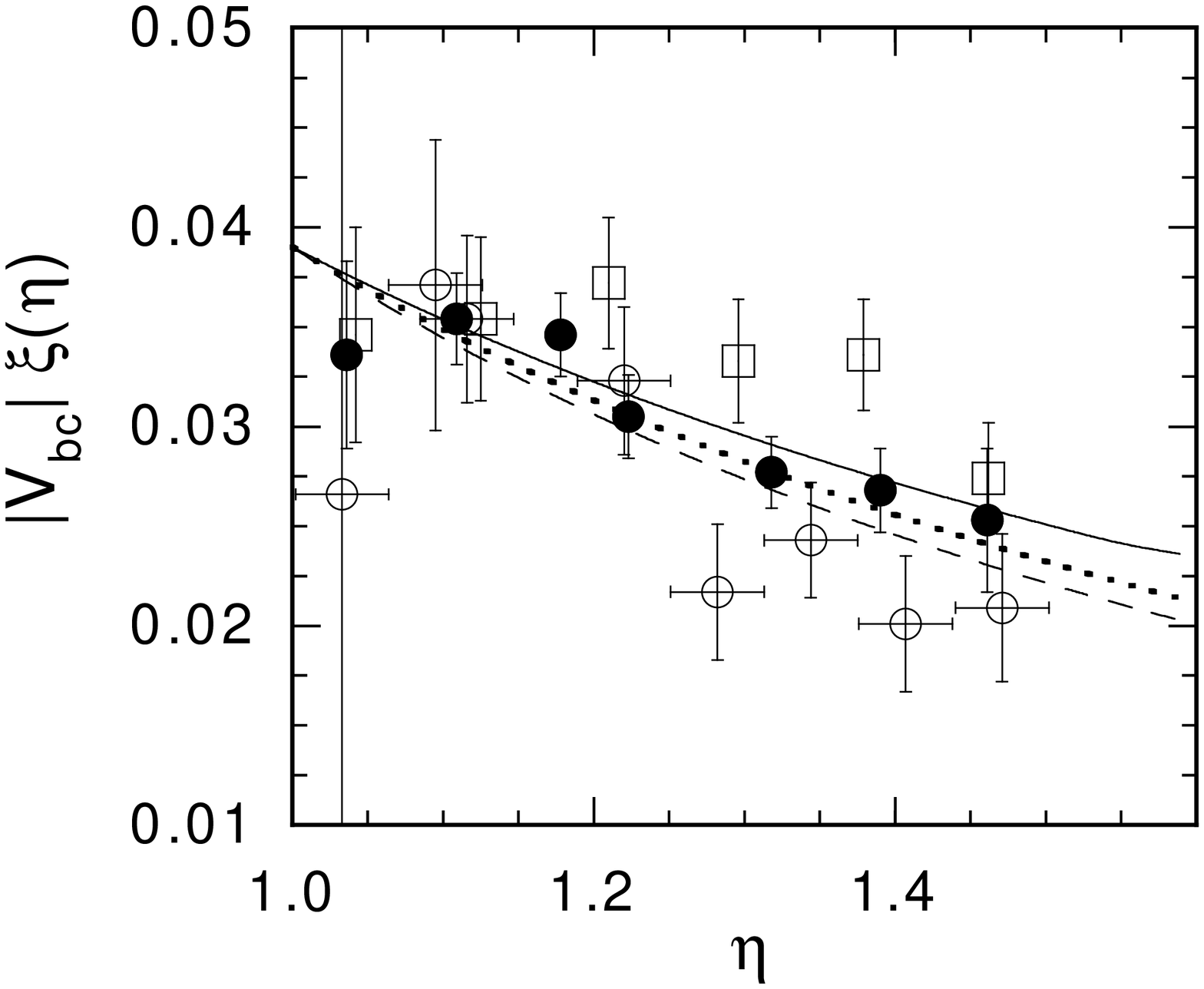}}
\vspace{-7cm}
\caption{The $IW$ form factor $\xi(\eta)$, times $|V_{bc}|$, as a 
function of the velocity recoil $\eta$. The open dots, full dots and 
squares correspond to the experimental data of Refs. [12], [13] and 
[14], respectively. The dotted, dashed and solid lines are our results 
obtained assuming $|V_{bc}| = 0.0390$ [6,7] and adopting the wave 
functions $A$, $B$ and $C$, respectively (see text). Taken from Ref. 
[7].}
\label{fig:fig1}
\end{center}
\end{figure}

\indent As for the exclusive $B \to D \ell \nu_{\ell}$ and 
$B \to D^* \ell \nu_{\ell}$ channels, the heavy quark limit of the 
relevant transition form factors is assumed and, therefore, the 
differential decay rates can be given only in terms of the Isgur-Wise 
($IW$) function \cite{IW89}. The latter has been explicitly calculated 
in case of the wave functions $A$, $B$ and $C$ in Refs. \cite{MT96}, 
\cite{SIM96} and \cite{GNST}, respectively. The results are reported 
in Fig. \ref{fig:fig1} and compared with the experimental data of Refs. 
\cite{ARGUS,CLEO,ALEPH}. It can clearly be seen that in the accessible 
recoil range our (model-dependent) predictions nicely reproduce 
exclusive $SL$ experimental data. It is worth noting that the calculated 
slope ($\rho^2$) of the $IW$ form factor turns out to be $\sim 1$ for 
all the three wave functions, i.e. $\sim 30\%$ higher than the prediction 
of the $ISGW$ quark model \cite{ISGW} ($\rho_{ISGW}^2 \sim 0.74$).

\begin{figure}[htb]
\begin{center}
\vspace{0.5cm}
\parbox{12cm}{\epsfxsize=12cm \epsfbox{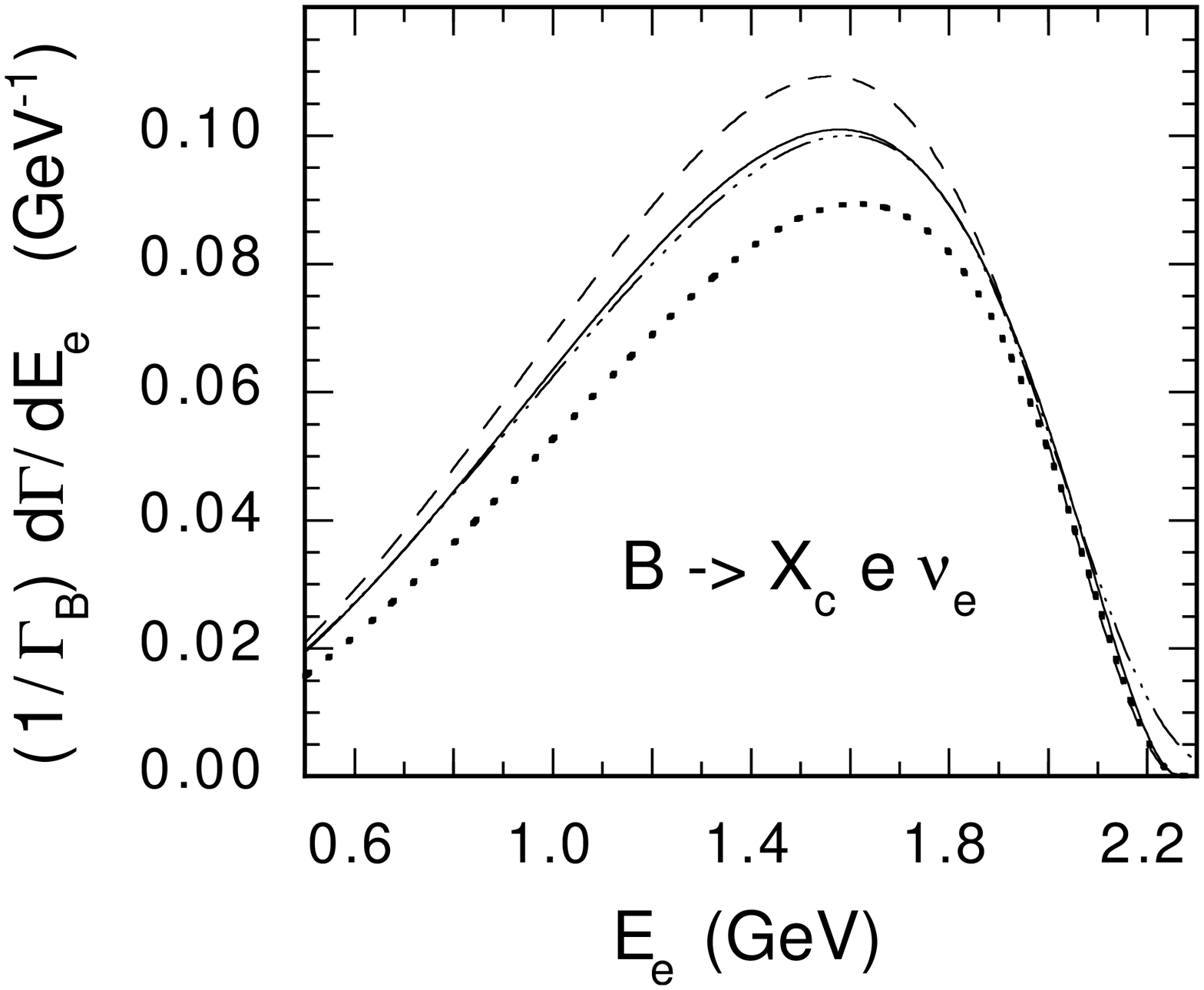}}
\vspace{-7cm}
\caption{The electron spectrum ${1 \over \Gamma_B} {d\Gamma \over dE_e}$
as a function of the electron energy $E_e$ for the inclusive 
$B \to X_c e \nu_e$ decays. The dot-dashed line is the result obtained 
within the convolution approach of Ref. [19]. The meaning of the dotted, 
dashed and solid lines is the same as in Figure 1.}
\label{fig:fig2}
\end{center}
\end{figure}

\indent As for the inclusive $SL$ width, our approach, fully described 
in Refs. \cite{MT96,GNST}, is essentially inspired to the deep inelastic 
scattering approach \cite{JPP94}, which pictures the heavy-meson decay 
as the decay of its partons. Our basic approximations for the calculation 
of the relevant weak hadronic tensor are the replacement of: ~ i) the 
matrix element $\langle n| J_{\mu}^{(h)} | B \rangle$ of the weak hadron 
current $J_{\mu}^{(h)}$ between the B-meson state and the final 
multi-hadron state $| n \rangle$ with the matrix element 
$\langle c | J_{\mu}^{(q)}  | b \rangle$ of the elementary 
$b \to c W^-$ weak transition times the non-perturbative $B$-meson 
wave function $\chi(x, \vec{p}_{\perp})$; ~ ii) the multi-hadron phase 
space $d\tau_n$ with the two-body phase space $d\tau_2(p_c, p_{sp})$, 
where $p_c$ and $p_{sp}$ are the final $c$-quark and spectator quark 
momenta. Our results for various $SL$ branching ratios are collected in 
Table \ref{tab:table1}, while the electron spectrum for the inclusive 
$B \to X_c e \nu_e$ process is shown in Fig. \ref{fig:fig2} and 
compared with the predictions of the convolution approach of Ref. 
\cite{MN94}, based on a partial resummation of the $OPE$ in the end-point 
region. It can clearly be seen that our predictions are in nice agreement
with the experimental $SL$ branching ratios and, at the same time, our 
calculated electron spectrum is consistent with the partially resummed 
$OPE$ result.

\begin{table*}[htb]
\vspace{0.5cm}
\setlength{\tabcolsep}{1.5pc}
\newlength{\digitwidth} \settowidth{\digitwidth}{\rm 0}
\catcode`?=\active \def?{\kern\digitwidth}
\caption{Branching ratios for semi-leptonic charmed decays of the 
$B$-meson. Cases $A$, $B$ and $C$ are described in the text. Equal 
charged and neutral $B$-meson production rates are assumed for the 
experimental data.}
\label{tab:table1}
\begin{tabular*}{\textwidth}{@{}l@{\extracolsep{\fill}}rrrr}
\hline
Decay mode & Case A & Case B & Case C& Exp. data\\
\hline
${\cal{B}}r(X_c \ell \nu_{\ell})$ & $10.06$ & $11.59$ & $12.32$
                                  & $10.77 \pm 0.43?{\cite{SKW}}$\\
${\cal{B}}r(X_c \tau \nu_{\tau})$ & $ 2.26$ & $ 2.46$ & $ 2.56$
                                  & $?2.60 \pm 0.32?{\cite{SKW}}$\\
${\cal{B}}r(D e \nu_e)$           & $ 1.83$ & $ 2.07$ & $ 1.69$
                                  & $?1.75 \pm 0.43?{\cite{PDG}}$\\
${\cal{B}}r(D^* e \nu_e)$         & $ 5.54$ & $ 5.98$ & $ 5.26$
                                  & $?4.93 \pm 0.42?{\cite{PDG}}$\\
\hline
Adapted from Ref. \cite{GNST}.
\end{tabular*}

\end{table*}

\begin{figure}[htb]
\begin{center}
\vspace{0.5cm}
\parbox{12cm}{\epsfxsize=12cm \epsfbox{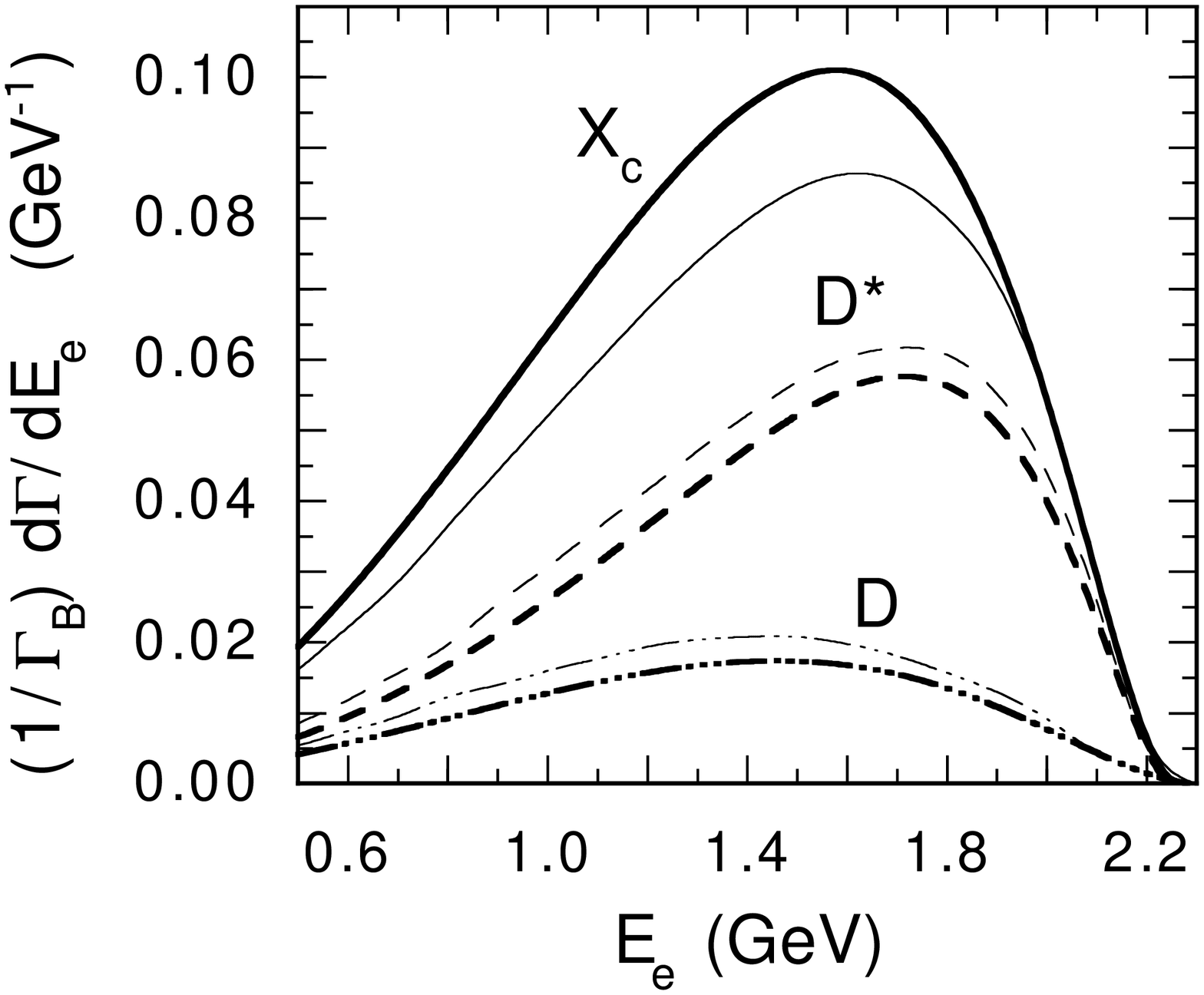}}
\vspace{-7cm}
\caption{The electron spectrum ${1 \over \Gamma_B} {d\Gamma \over dE_e}$
as a function of the electron energy $E_e$ for the "resonant" 
$B \to D e \nu_e$ (dotted lines) and $B \to D^* e \nu_e$ (dashed lines) 
channels, as well as for the total inclusive $B \to X_c e \nu_e$ decay 
(solid lines). Thick lines correspond to the results obtained in our 
case $B$, whereas thin lines are the predictions of the $ISGW$ model 
[15].}
\label{fig:fig3}
\end{center}
\end{figure}

From Table \ref{tab:table1} it can also be seen that our results for the
contribution of the "resonant" $B \to D \ell \nu_{\ell}$ and 
$B \to D^* \ell \nu_{\ell}$ channels to the total $SL$ branching ratio 
remarkably differ from the predictions of the $ISGW$ quark model 
\cite{ISGW}. At variance with this model (and in accord with the 
experimental data), we find that $\sim 20 \%$ and $\sim 70 \%$ of the 
total $SL$ rate is due to the exclusive $D$ and $D + D^*$ decay modes, 
respectively, leaving a remarkable fraction ($\sim 30 \%$) of the 
strength to other final states. Moreover, in Fig. \ref{fig:fig3} our 
results for the electron spectra are compared with the corresponding 
$ISGW$ prediction \cite{ISGW}. It can be seen that, with respect to the 
results of our quark model, the $ISGW$ model predicts both a larger 
strength in the resonant $D$ and $D^*$ channels and a lower strength in 
the total inclusive spectrum.

\indent The results so far presented clearly illustrate that our 
quark model is successfull in describing bound-state effects both 
in exclusive and inclusive $SL$ charmed decays. Therefore, we have 
applied our approach to the calculation of the inclusive widths of $NL$ 
charmed decays (see Refs. \cite{MT96,GNST}). For these processes our 
basic assumption is to incorporate all the long-range $QCD$ effects due 
to soft gluons into our non-perturbative wave function of the initial 
$B$-meson and the final $D$ or $D^*$ mesons. Only the corrections due to 
hard-gluon exchange, yielding the effective weak Lagrangian of Ref. 
\cite{VZS79}, are taken into account. Some of our results are collected 
in Table \ref{tab:table2} and compared with experimental data. It can be 
seen that a good agreement (within $2 \sigma$) is achieved and, in 
particular, a non-negligible fraction of decays into baryon-antibaryon 
pairs is found. Note the good agreement of our predicted charm counting 
with recent experimental findings. Before closing, we want to point out 
that: ~ i) the sum of all the calculated branching ratios turns out to 
be very close to $1$ (see \cite{GNST}); ~ ii) radiative $QCD$ corrections 
have been so far neglected; their net effect is expected to be of the 
same order of the uncertainties related to the model dependence of the 
$B$-meson wave function.

\begin{table*}[htb]
\vspace{0.5cm}
\setlength{\tabcolsep}{1.5pc}
\settowidth{\digitwidth}{\rm 0}
\catcode`?=\active \def?{\kern\digitwidth}
\caption{Branching ratios for some non-leptonic charmed decays of the 
$B$-meson. Cases $A$, $B$ and $C$ are as in Table 1. Equal charged and 
neutral $B$-meson production rates are assumed for the experimental 
data.}
\label{tab:table2}
\begin{tabular*}{\textwidth}{@{}l@{\extracolsep{\fill}}rrrr}
\hline
Decay mode & Case A & Case B & Case C& Exp. data\\
\hline
${\cal{B}}r(D \pi)$             & $0.40$ & $0.49$ & $0.36$
                                & $0.41 \pm 0.03?{\cite{PDG}}$\\
${\cal{B}}r(D \rho)$            & $0.90$ & $1.08$ & $0.82$
                                & $1.06 \pm 0.11?{\cite{PDG}}$\\
${\cal{B}}r(D^* \pi)$           & $0.42$ & $0.49$ & $0.38$
                                & $0.39 \pm 0.04?{\cite{PDG}}$\\
${\cal{B}}r(D^* \rho)$          & $1.08$ & $1.25$ & $0.99$
                                & $1.14 \pm 0.17?{\cite{PDG}}$\\
${\cal{B}}r(\bar{D} D_s)$       & $1.55$ & $1.75$ & $1.43$
                                & $1.10 \pm 0.35?{\cite{CLEO_D}}$\\
${\cal{B}}r(\bar{D} D_s^*)$     & $1.28$ & $1.43$ & $1.19$
                                & $0.89 \pm 0.31?{\cite{CLEO_D}}$\\
${\cal{B}}r(\bar{D}^* D_s)$     & $1.09$ & $1.20$ & $1.02$
                                & $1.12 \pm 0.36?{\cite{CLEO_D}}$\\
${\cal{B}}r(\bar{D}^* D_s^*)$   & $3.35$ & $3.66$ & $3.15$
                                & $2.41 \pm 0.74?{\cite{CLEO_D}}$\\
${\cal{B}}r(charmed ~ baryons)$ & $8.24$ & $5.15$ & $3.86$
                                & $6.4? \pm 1.1??{\cite{PDG}}$\\
$charm ~ counting$              & $1.21$ & $1.20$ & $1.20$
                                & $1.16 \pm 0.05?{\cite{SKW}}$\\
                          & & & & $1.23 \pm 0.07?{\cite{ALEPH_C}}$\\
\hline
Adapted from Ref. \cite{GNST}.
\end{tabular*}

\end{table*}

\section{CONCLUSIONS}

A phenomenological quark model, which takes into account non-perturbative 
$QCD$ bound-state effects, has been applied to the investigation of 
exclusive and inclusive, semileptonic and non-leptonic, charmed decays 
of the $B$-meson, obtaining a remarkable agreement with available 
experimental data.


\vspace{1cm}

\begin{thebibliography}{99}

\bibitem{HQET} For a review see, e.g., M. Neubert, Phys. Rep. 245 (1994) 
 259, and references therein quoted.

\bibitem{OPE} I.I. Bigi, N.G. Uraltsev and A.I. Vainshtein, Phys. Lett. 
 B247 (1992) 293. ~ I.I. Bigi, M.A. Shifman, N.G. Uraltsev and A.I. 
 Vainshtein, Phys. Rev. Lett. 71 (1993) 496; Int. J. Mod. Phys. A9 (1994) 
 2647.  ~ A.V. Manohar and M.B. Wise, Phys. Rev. D49 (1994) 1310. ~ A.F. 
 Falk, M. Luke and M.J. Savage, Phys. Rev. D49 (1994) 3367. ~ M. Neubert, 
 Phys. Rev. D49 (1994) 3392 and 4623. ~ B. Block, L. Koyrakh, M.A. 
 Shifman and A.I. Vainshtein, Phys. Rev. D50 (1994) 3356. ~ T. Mannel, 
 Nucl. Phys. B413 (1994) 396.

\bibitem{PQW76} E.C. Poggio, H.R. Quinn and S. Weinberg, Phys. Rev. D13 
 (1976) 1958.

\bibitem{GN96} M. Girone and M. Neubert, Phys. Rev. Lett. 76 (1996) 3061.

\bibitem{AMPR96} G. Altarelli, G. Martinelli, S. Petrarca and F. Rapuano,
 preprint CERN-TH/96-77.

\bibitem{MT96} V.L. Morgunov and K.A. Ter-Martirosyan, Physics of Atomic
 Nuclei 59 (1996) 1221.

\bibitem{GNST} I.L. Grach, I.M. Narodetskii, S. Simula and K.A.
 Ter-Martirosyan, hep-ph 9603329.

\bibitem{GI85} S. Godfrey and N. Isgur, Phys. Rev. D32 (1985) 185.

\bibitem{NCS92} I.M. Narodetskii, R. Ceuleener and C. Semay, J. Phys.
 G18 (1992) 1901.

\bibitem{IW89} N. Isgur and M.B. Wise, Phys. Lett. B232 (1989) 113 
 and B237 (1990) 447.

\bibitem{SIM96} S. Simula, Phys. Lett. B373 (1996) 193.

\bibitem{ARGUS} ARGUS coll., H. Albrecht et al., Z. Phys. C57 (1993) 
 533.

\bibitem{CLEO} CLEO coll., B. Barish et al., Phys. Rev. D51 (1995) 
 1014.

\bibitem{ALEPH} ALEPH coll., D. Buskulic et al., Phys. Lett. B359 
 (1995) 236.

\bibitem{ISGW} For the updated version of the $ISGW$ model see D. 
 Scora and N. Isgur, Phys. Rev. D52 (1995) 2783.

\bibitem{JPP94} See, e.g., C.H. Jin, M.F. Palmer and E.A. Paschos, 
 Phys. Lett. B329 (1994) 364, and references therein quoted.

\bibitem{SKW} T. Skwarnicki, in Proc. of the $17^{th}$ Int. Symp. on 
 Lepton-Photon Interactions, Beijing (China), August 1995, eds. Z. 
 Zhi-Peng and C. He-Sheng (World Scientific, Singapore, 1996), p. 
 238.

\bibitem{PDG} Particle Data Group, R.M. Barnett et al.: Phys. Rev. 
 D53 (1996) 1.

\bibitem{MN94} T. Mannel and M. Neubert, Phys. Rev. D50 (1994) 2037.

\bibitem{VZS79} A.I. Vainstein, V.I. Zakharov and M.A. Shifman, Nucl. 
 Phys. B147 (1979) 385 and 448.

\bibitem{CLEO_D} CLEO coll., D. Gibaut et al., Phys. Rev. D53 (1996) 
 4734.

\bibitem{ALEPH_C} ALEPH coll., D. Buskulic et al., preprint 
 CERN-PPE/96-117.

\end{thebibliography}
\end{document}